# Experimental Progress on the Emergent Infinite-Layer Ni-Based Superconductors

*Xiaorong Zhou[#], Peixin Qin[#], Zexin Feng[#], Han Yan, Xiaoning Wang, Hongyu Chen, Ziang Meng, and Zhiqi Liu\**

X. Zhou, P. Qin, Z. Feng, H. Yan, X. Wang, H. Chen, Z. Meng, Prof. Z. Liu
School of Materials Science and Engineering, Beihang University, Beijing 100191, China
E-mail: zhiqi@buaa.edu.cn

[#]These authors contributed equally to this work.

**The emergence of the infinite-layer superconducting nickelate thin films marks the Ni age of superconductivity, which has excited a huge surge of studies since the first report in August of 2019. Despite of the tremendous attention drawn from the entire material science community and a large body of theoretical studies, the experimental progress has been relatively slow due to the challenging sample fabrication, which may, in turn, be holding back the fast development of theoretical research. Therefore, a timely and comprehensive review on all the up-to-date experimental progress of the emergent infinite-layer Ni-based superconductors is urgently needed. In this review, we first introduce the history of more than 30-year-long Ni-based superconductivity exploration, then summarize the sample fabrication processes, later present the experimental electrical transport and magnetic properties, and finally come up with several key issues deserving intensive studies. This review is thus expected to be helpful for researchers with diverse research background to readily capture the major progress of this emerging field.**

# 1. Introduction

Since the discovery of high-temperature superconducting cuprates in 1986 [1], possible new superconductors with similar electronic and crystal structures have been pursued for decades. Nickel-based compounds may be promising candidates for achieving superconductivity because of the proximity of Ni and Cu in the periodic table. It is known that the +2 valence is the most usual valence state of Ni ions. However, no superconducting state was found in compounds with $Ni^{2+}$ ions in the 20th century. Despite of early claims on the superconductivity in $La_2NiO_4$ systems based on purely magnetic-moment measurements in 1989 [2-4], reliable transport data on superconductivity were not achieved [5]. Almost twenty years later, with the advent of Fe-based superconductors, a series of layered nickel pnictides or chalcogenide compounds exhibiting superconducting transition with predominant $Ni^{2+}$ ions were successfully synthesized such as LaNiOP [6], LaNiOAs [7], $BaNi_2P_2$ [8], $BaNi_2As_2$ [9], $La_3Ni_4P_4O_2$ [10], $LaNiAsO_{1-x}H_x$ [11] and other compounds [12]. Nevertheless, due to the relatively low $T_c$ (below 5 K) of these compounds, this type of superconductors could hardly beat the Fe-based superconductors with similar structures. As for $Ni^{3+}$ compounds, in 2008, Chaloupka et al. [13] predicted that superconductivity may exist in $LaNiO_3$/$LaMO_3$ ($M$ represents transition metal elements) superlattices owing to its nondegenerate spin one-half electronic structure akin to cuprates. Thus, great efforts were made to probe this type of superlattices [14-17]. Some promising experimental results have been reported by Zhou et al [18], where both the superconductivity transition occurring at 3.7 K and the Meissner effect were observed in a $LaNiO_3$/$La_{0.7}Sr_{0.3}MnO_3$ superlattice.

Apart from the +2 and +3 valence states of Ni, $Ni^+$ possesses the same $3d^9$ electronic configuration as $Cu^{2+}$ in cuprates. Indeed, as early as in 1999, Anisimov et al. [19] proposed theoretically that $LaNiO_2$ may act as an analog of a cuprate. Despite many experimental realization of $Ni^+$ oxides via chemical hydrogen reduction in both the bulk [20-25] and thin-film [26-29] forms such as $LaNiO_2$ [20,21,27-29], $NdNiO_2$ [22,26], $R_4Ni_3O_8$ [23,25], $La_3Ni_2O_6$ [24], the superconductivity had not been realized in Ni$^+$ compounds. Until the August of 2019, Li et al. [30] reported the successful observation of superconductivity with onset temperatures of 9-15 K in infinite-layer Nd$_{0.8}$Sr$_{0.2}$NiO$_2$ thin films, which instantly attracted intense research interest among the broad materials science community. Inspired by this breakthrough, a huge surge of studies on the emergent infinite-layer Ni-based superconductors has soon showed up. According to Web of Science, the first report [30] has been cited by more than 250 times (up to February of 2022), and the number of relevant follow-up papers is increased by 20-30 every three months in the past two years (Fig. 1).

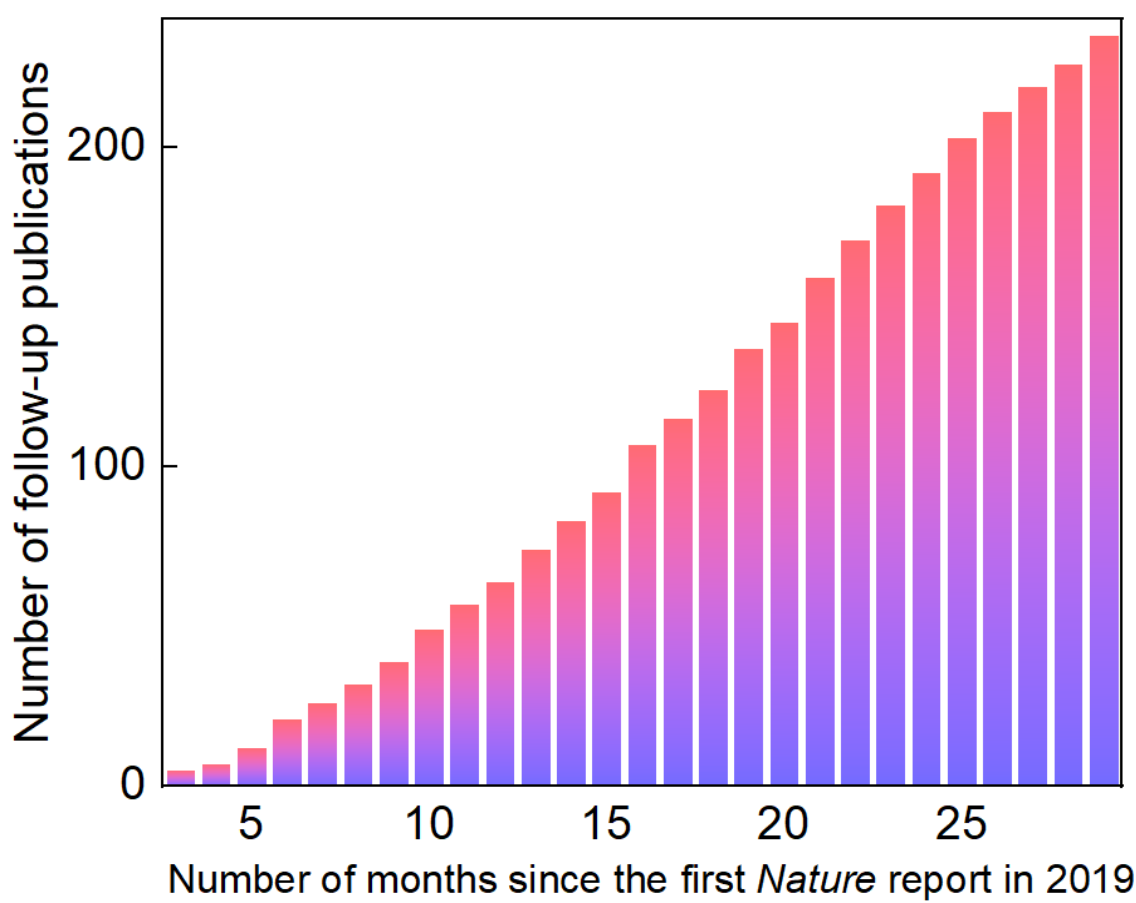

**FIGURE 1** Accumulative number of publications (up to February of 2022) on the emergent infinite-layer Ni-based superconductors indexed in Web of Science since the first report in August of 2019.

Meanwhile, a number of research highlights and reviews [31-37] have been published to emphasize the scientific significance of the superconductivity discovered in infinite-layer nickelates. For example, in 2019, George A. Sawatzky in University of British Columbia wrote a New & View paper [31] entitled "Superconductivity seen in a non-magnetic nickel oxide"; a news report [32] highlighted the work with the title of "Superconductivity is found in a nickel oxide". In 2020, Michael R. Norman at Argonne National Laboratory composed a trend paper [33] entitled "Entering the nickel age of superconductivity". In 2021, Warren E. Pickett in University of California Davis picked up superconductivity in infinite-layer nickelates as one of the research highlights of 2020 and wrote a paper [34] with the subject of "The dawn of the nickel age of superconductivity".

Although a large number of theoretical studies [38-87] have been undertaken on this new superconducting system, the experimental works achieving superconductivity are just a tiny minority in the large body of the follow-up studies. To our best knowledge, there have been only five independent research groups claiming the successful synthesis of superconducting infinite-layer nickelates thin films. Specifically, most of experimental works were contributed by Harold Hwang's group at Stanford University who discovered the superconductivity of infinite-layer nickelate thin films [30, 88-97], and other experimental works came from Hai-Hu Wen's group at Nanjing University [98-100], Ariando's group at National University of Singapore [101,102], Xingjiang Zhou's group at Institute of Physics, Chinese Academy of Sciences [103-105], and recently our own group [106,107].

The main reasons for the rather limited successful sample fabrication could possibly be: (1) both the quality of 113-phase $Nd_{0.8}Sr_{0.2}NiO_3$ films and the chemical reduction process seem to be critical but it is not clear yet what factors of the 113-phase films are crucial; (2) there are so many parameters in the chemical reduction process to be optimized such as the amount of the reduction agent - $CaH_2$ powder, assisting temperature, chemical reaction time, the volume of Pyrex tubes, the thin-film thickness versus all these chemical reduction processes and even the relative positions of thin-film samples and the $CaH_2$ powder. Consequently, most of sample fabrication processes are rather blind and whether one can finally achieve superconductivity is highly random. It is kind of embarrassing that one has to attempt for many samples over and over again until a superconducting sample with proper structure appears. Also because of the poor control of the multiple factors for the sample preparation, the nominal same fabrication process could yield reduced samples with completely opposite transport properties.

Due to the lack of sufficient experimental progress, the theoretical studies might thus be performed in a rather scattered manner, and many theoretical results do not agree with each other and might even be contrary to later experimental results. On the other hand, though the experimental reports are rare, most of them are consistent with each other up to now. Therefore, it is of immediate need to constructively summarize the experimental aspects to promote the rapid development of the entire field. In this article, all the up-to-date experimental works including the synthesis, electrical and magnetic properties of nickelates superconductor are comprehensively reviewed, and some outstanding key issues from the experimental perspective are also emphasized, which may be critical for this emergent field.

## 2. Synthesis of nickelates

### 2.1 Failed synthesis without chemical reduction

A superconducting nickelate $Nd_{0.8}Sr_{0.2}NiO_2$ is an oxygen-poor phase which is reduced from a usual $Nd_{0.8}Sr_{0.2}NiO_3$ compound. The chemical reduction reaction often utilizes $CaH_2$ powder as a reductant and needs a series of complicated processes. Therefore, an alternative non-chemical reduction approach is highly desirable to simplify the whole reduction process. Motivated by this idea, our group fabricated various $Nd_{0.8}Sr_{0.2}NiO_x$ thin films via modulating the oxygen partial pressure during the deposition process in a pulsed laser deposition chamber [108]. Seven different oxygen pressures and nine types of single oxide substrates were utilized to synthesize 63 types of $Nd_{0.8}Sr_{0.2}NiO_x$ thin films. X-ray diffraction (Fig. 2a) shows that, with high oxygen pressures of 150 and 100 mTorr, the single-crystal $Nd_{0.8}Sr_{0.2}NiO_3$ phase was formed with a clear (002) peak. However, no single-crystalline $Nd_{0.8}Sr_{0.2}NiO_3$ phase was found with lower oxygen pressures that were expected to reduce the $Nd_{0.8}Sr_{0.2}NiO_3$ films more effectively. This agrees with the cross-section transmission electron microscope image (Fig.

2b) of $Nd_{0.8}Sr_{0.2}NiO_x$ thin films deposited with an oxygen pressure of $10^{-6}$ Torr, which reveals amorphous and nanocrystalline nature of the interface regions.

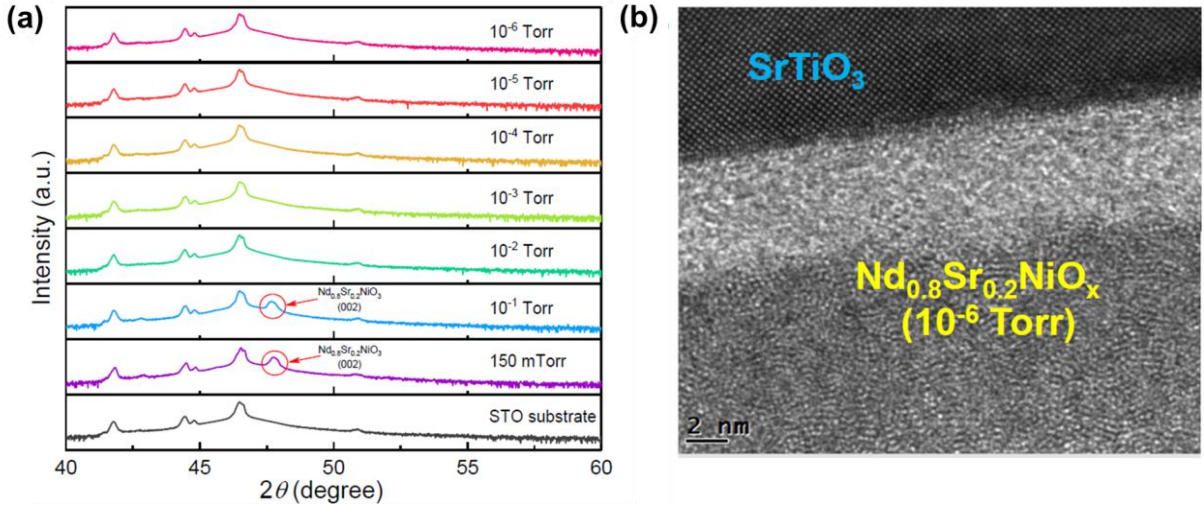

**FIGURE 2** (a) X-ray diffraction pattern of a $SrTiO_3$ substrate and $Nd_{0.8}Sr_{0.2}NiO_x/SrTiO_3$ heterostructures with different oxygen partial pressures. (b) Cross-section transmission electron microscopy image of a $Nd_{0.8}Sr_{0.2}NiO_x/SrTiO_3$ heterostructure fabricated with an oxygen partial pressure of $10^{-6}$ Torr. Reproduced with permission from Ref. [108]. Copyright 2020, Springer Nature.

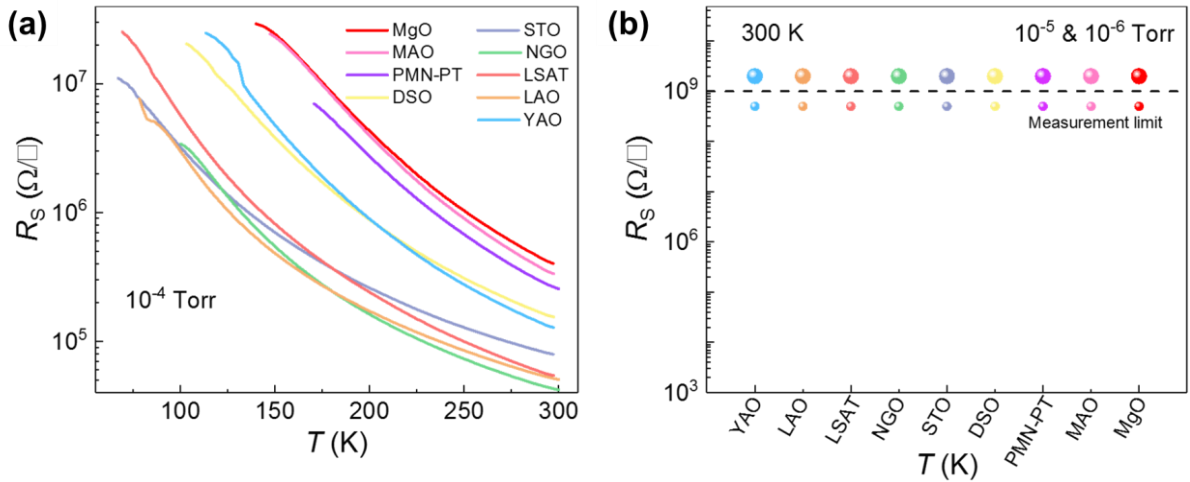

**FIGURE 3** Sheet resistance of $Nd_{0.8}Sr_{0.2}NiO_x$ thin films deposited onto different substrates with low oxygen partial pressures (a) $10^{-4}$ Torr, (b) $10^{-5}$ & $10^{-6}$ Torr at different temperatures. Reproduced with permission from Ref. [108]. Copyright 2020, Springer Nature.

According to our electrical transport measurements (Fig. 3), superconductivity was not found in all these 63 types of samples, and with lowering the oxygen pressure, the $Nd_{0.8}Sr_{0.2}NiO_x$ thin films became more and more resistive and finally turned to be highly insulating. This suggests that the chemical reduction process using $CaH_2$ could be an indispensable approach to generating the 112-phase superconducting infinite-layer nickelates. It is worthy noticing that Alex Zunger's recent theoretic study [39] may give a plausible explanation on the absence of superconductivity in $Nd_{0.8}Sr_{0.2}NiO_x$ thin films without chemical reduction. In this work, using density functional theory calculations, they found that the existence of H atoms from in the chemical reduction reaction could be vital to stabilize the $NdNiO_2$ phase via forming a hydride compound $NdNiO_2H$.

## 2.2 Synthesis with chemical reduction using vacuum tubes

The soft chemical topotactic reduction is hitherto the only way successfully converting $Nd_{1-x}Sr_xNiO_3$ to superconducting $Nd_{1-x}Sr_xNiO_2$. In general, this kind of solid reduction reaction requires $CaH_2$ powder as a reductant, aluminum foil-wrapped $Nd_{1-x}Sr_xNiO_3$ samples and vacuum environment such as sealed glass or quartz tubes. For example, in 2019, Li et al. [30] from Harold Hwang's group at Stanford University firstly reported the successful synthesis of superconducting $Nd_{0.8}Sr_{0.2}NiO_2$ thin films using this method. Firstly, $Nd_{0.8}Sr_{0.2}NiO_3$ films were deposited onto $TiO_2$-terminated $SrTiO_3$ substrates by pulsed laser deposition (PLD) with epitaxial $SrTiO_3$ capping layers. Then the 2.5×5 mm$^2$ $Nd_{0.8}Sr_{0.2}NiO_3$ samples loosely wrapped by aluminum foils were vacuum-sealed together with $CaH_2$ powder in a Pyrex glass tube and the tube was heated to 260-280°C for 4-6 h. Finally, 9-11 nm superconducting nickelate thin films were obtained. X-ray diffraction (Fig. 4a) presents clear single-crystal (001) and (002) peaks of $Nd_{0.8}Sr_{0.2}NiO_2$. Temperature-dependent resistivity (Fig. 4b) shows a superconducting transition occurring at an onset temperature of 14.9 K for a hole-doped $Nd_{0.8}Sr_{0.2}NiO_2/SrTiO_3$ thin film.

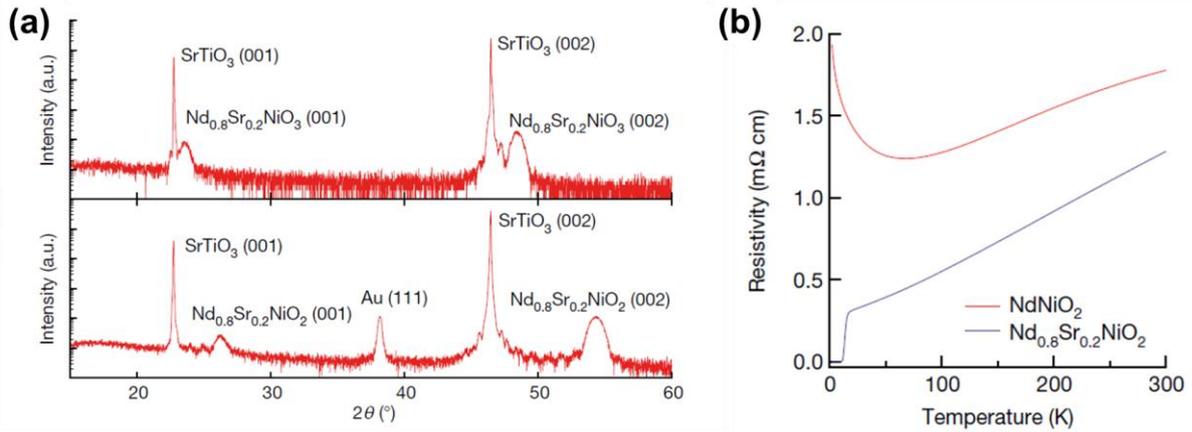

**FIGURE 4** (a) X-ray diffraction pattern of a $Nd_{0.8}Sr_{0.2}NiO_3/SrTiO_3$ and a $Nd_{0.8}Sr_{0.2}NiO_2/SrTiO_3$ heterostructures. (b) Resistivity of the $NdNiO_2$ and $Nd_{0.8}Sr_{0.2}NiO_2$ thin films as a function of temperature. Reproduced with permission from Ref. [30]. Copyright 2019, Springer Nature.

In 2020, Gu et al. [98] from Hai-Hu Wen's group at Nanjing University successfully prepared superconducting $Nd_{0.8}Sr_{0.2}NiO_2$ thin films via the similar procedure. They grew 6-nm-thick $Nd_{0.8}Sr_{0.2}NiO_3$ thin films without $SrTiO_3$ capping layers using reactive molecular beam epitaxy. The obtained $Nd_{0.8}Sr_{0.2}NiO_3$ thin films were then placed in an evacuated quartz tube with $CaH_2$ power and finally the tube was heated to 260-280°C for 4-6 h. Fig. 5 plots temperature-dependent resistivity of a reduced thin-film sample. The onset transition temperature is 15.3 K and the zero resistivity arises at 9.1 K.

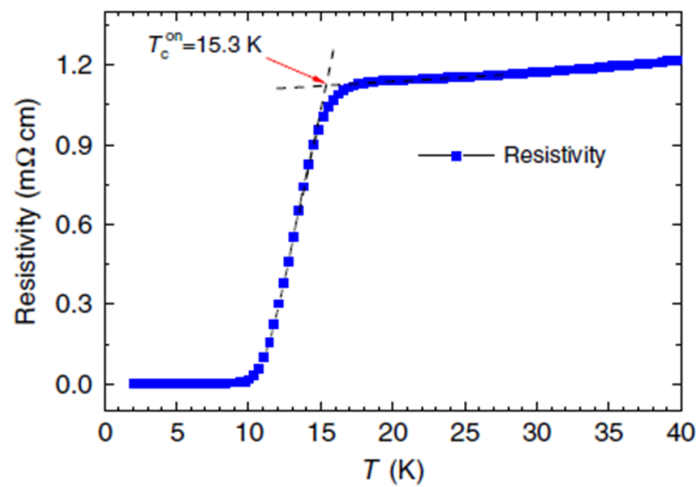

**FIGURE 5** Resistivity of a $Nd_{0.8}Sr_{0.2}NiO_2/SrTiO_3$ film as a function of temperature. Reproduced with permission from Ref. [98]. Copyright 2020, Springer Nature.

In 2021, Gao et al. [105] from Xingjiang Zhou's group at Institute of Physics, Chinese Academy of Sciences demonstrated the fabrication of superconducting $Nd_{0.8}Sr_{0.2}NiO_2$ thin films by chemical reduction with a glass tube. The procedure in this work is almost the same with that of Ref. [26]. As shown in Fig. 6a-c, X ray diffraction confirms the single-crystal phase of $Nd_{0.8}Sr_{0.2}NiO_2$ and an onset superconducting transition temperature of 9 K was achieved for a thin-film sample.

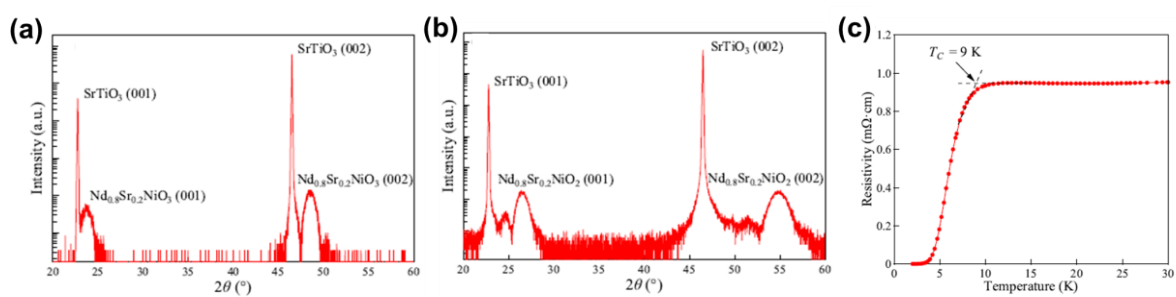

**FIGURE 6** (a) X-ray diffraction pattern of a $Nd_{0.8}Sr_{0.2}NiO_3/SrTiO_3$ heterostructure. (b) X-ray diffraction pattern of a $Nd_{0.8}Sr_{0.2}NiO_2/SrTiO_3$ heterostructure. (c) Temperature-dependent resistivity of the $Nd_{0.8}Sr_{0.2}NiO_2/SrTiO_3$ sample. Reproduced with permission from Ref. [105]. Copyright 2021, IOP publishing.

Up to now, more experimental works [88-97] on achieving superconductivity in chemically reduced nickelates from these three groups have been published. Note that $Pr_{0.8}Sr_{0.2}NiO_2$ [91,92,103,104], $La_{1-x}Sr_xNiO_2$ [94] and undoped $Nd_6Ni_5O_{12}$ [96] that is reduced from its Ruddlesden-Popper parent phase $Nd_6Ni_5O_{16}$ have also been shown to be superconducting as well as $Nd_{0.8}Sr_{0.2}NiO_2$ thin films, which suggests the possible existence of a broad family of superconducting nickelates.

## 2.3 Synthesis with chemical reduction using PLD chambers

One of the major challenges in synthesizing $Nd_{1-x}Sr_xNiO_2$ thin films may come from the chemical reduction in vacuum-sealed tubes since most oxide electronics groups are working in the field of condensed matter physics and lack in relevant chemical experimental experience. In 2020, Zeng et al. [101,102] from Ariando's group at National University of Singapore adopted an alternative approach to preparing superconducting nickelate thin films. The obvious difference is that they used a PLD chamber instead of a vacuum glass tube to carry out the solid chemical reduction reactions. The pre-grown 35-nm-thick $Nd_{1-x}Sr_xNiO_3$ thin films embedded among $CaH_2$ powder were wrapped by aluminum foils and then placed in the PLD chamber. Then the chamber was heated to 340-360°C for 80-120 min. Clear (001) and (002) single-crystal peaks of $Nd_{1-x}Sr_xNiO_2$ can be seen for all the Sr-doping levels via X-ray diffraction patterns.

Fig. 7 shows temperature-dependent resistivity of $Nd_{1-x}Sr_xNiO_2$/$SrTiO_3$ thin films with different doping levels. The superconducting transition occurs in $Nd_{1-x}Sr_xNiO_2$ with a Sr doping range of $0.135 \leq x \leq 0.22$.

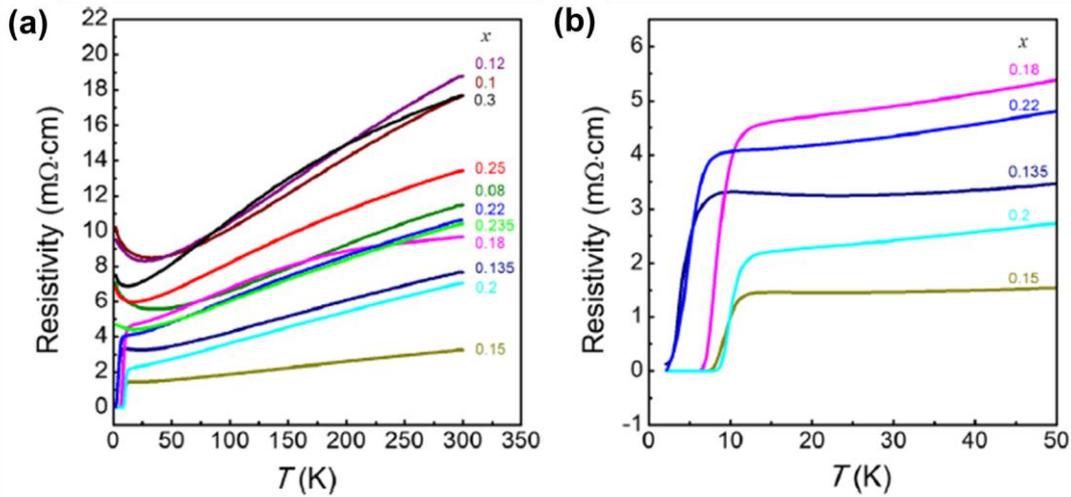

**FIGURE 7** (a) Temperature-dependent resistivity of $Nd_{1-x}Sr_xNiO_2$/$SrTiO_3$ heterostructures with different Sr-doping levels ranging from 0.08 to 0.3. (b) Zoom-in temperature dependence of resistivity of $Nd_{1-x}Sr_xNiO_2$/$SrTiO_3$ heterostructures with the Sr-doping levels ranging from 0.135 to 0.22. Reproduced with permission from Ref. [102]. Copyright 2020, American Physical Society.

Due to the convenience for oxide electronic groups that typically possess vacuum chambers capable of heating and relatively larger thickness of the obtained superconducting thin films with this method, our group has utilized this approach and successfully synthesized a series of superconducting $Nd_{0.8}Sr_{0.2}NiO_2$ thin films with different thickness ranging from 8 to 40 nm [106,107]. As shown in Fig. 8, the (001) and (002) peaks in the X-ray diffraction pattern locates at 26.5º and 54.6º, respectively, which agrees well with the previous report in Ref. [15]. The 360° phi scans assure the epitaxy of the $Nd_{0.8}Sr_{0.2}NiO_2$ thin film on the $SrTiO_3$ substrate. The X-ray reflectivity pattern indicates a smooth surface of the film. Fig. 9 presents the temperature-dependent resistivity of the $Nd_{0.8}Sr_{0.2}NiO_2$ thin film, which exhibits metallic behavior at relatively high temperatures and undergoes a superconducting transition at an onset temperature of ~13.4 K.

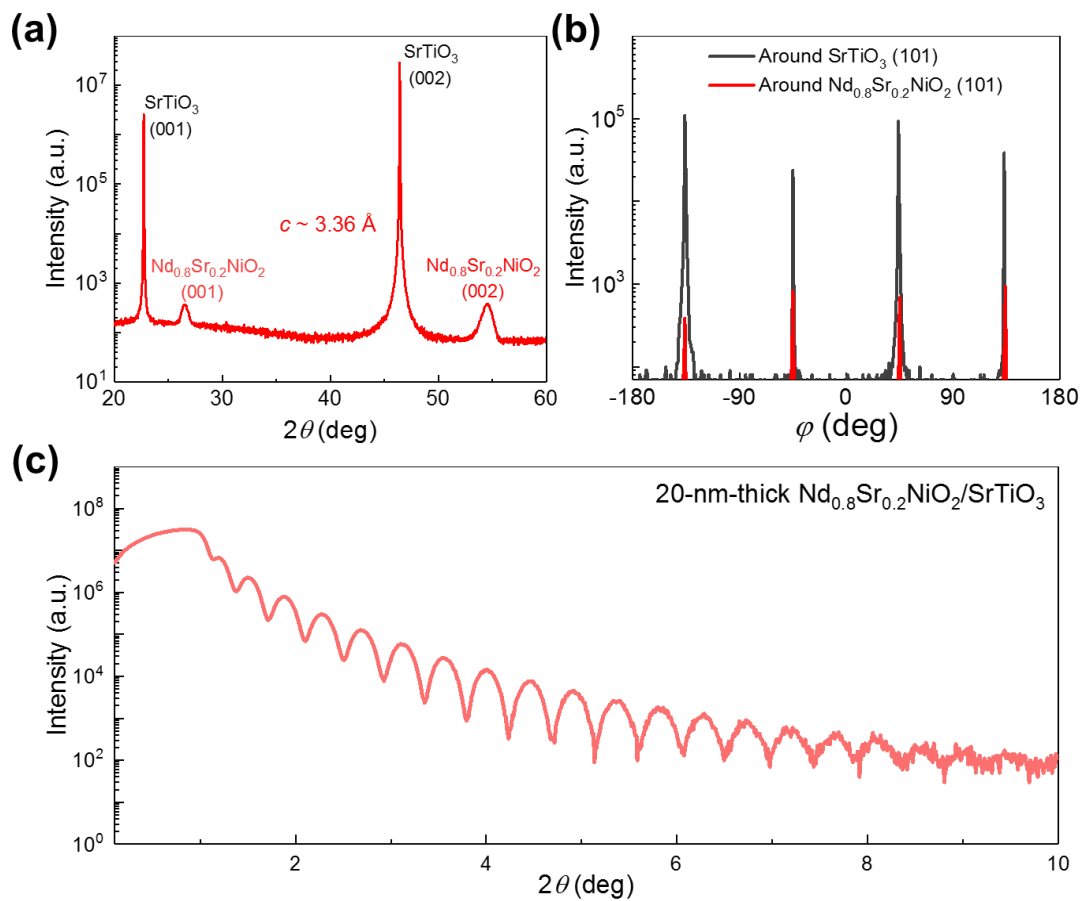

**FIGURE 8** (a) X-ray diffraction theta-2theta scan of a $Nd_{1-x}Sr_xNiO_2/SrTiO_3$ heterostructure. (b) 360° phi scans around the $Nd_{1-x}Sr_xNiO_2$ and $SrTiO_3$ (101) peaks. (c) X-ray reflectivity of the $Nd_{1-x}Sr_xNiO_2/SrTiO_3$ heterostructure. Reproduced with permission from Ref. [106]. Copyright 2021, Wiley-VCH.

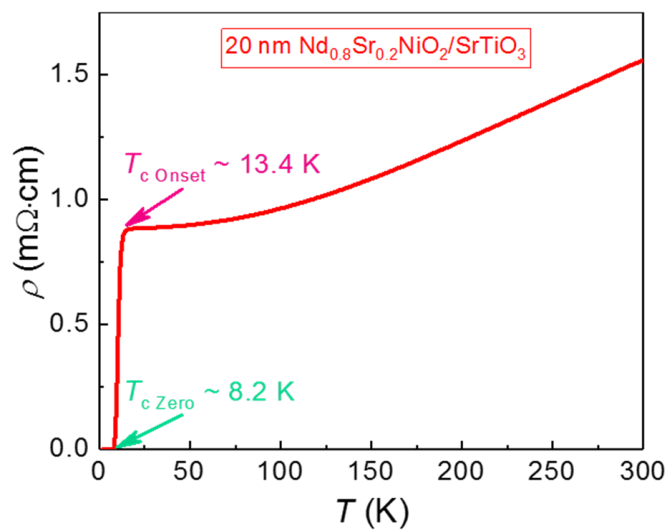

**FIGURE 9** Temperature dependence of resistivity of a 20-nm-thick Nd$_{1-x}$Sr$_x$NiO$_2$/SrTiO$_3$ heterostructure. Reproduced with permission from Ref. [106]. Copyright 2021, Wiley-VCH.

## 2.4 How about oxygen vacancies in SrTiO$_3$ substrates after chemical reduction?

Oxygen vacancies play important roles in oxides including catalysis [109-112], gas sensor applications [113], energy storage [114,115], electronic states [116] and in oxide interface systems [117-121] such as LaAlO$_3$/SrTiO$_3$ interfaces [122,123]. In the superconducting Nd$_{0.8}$Sr$_{0.2}$NiO$_2$/SrTiO$_3$ superconducting system, oxygen vacancies may be generated during the soft chemical reduction, which could contribute conducting channels and thus affect superconducting transitions. Therefore, it is necessary to examine to what degree oxygen vacancies are generated during the CaH$_2$ chemical reduction process. In our previous work [107], photoluminescence spectra were employed to qualitatively analyze the amount of oxygen vacancies. Multiple defect-level-related photoluminescence peaks between 350-600 nm are generated by oxygen vacancies and the intensities of the peaks are sensitive to the concentration of oxygen vacancies. Fig. 10 shows the photoluminescence of a bare SrTiO$_3$ crystal and a Nd$_{0.8}$Sr$_{0.2}$NiO$_2$/SrTiO$_3$ heterostructure excited by a 325 nm laser at room temperature. The intensity of photoluminescence is even weaker in the Nd$_{0.8}$Sr$_{0.2}$NiO$_2$/SrTiO$_3$ heterostructure compared to the bare SrTiO$_3$ crystal, which indicates that the concentration of oxygen vacancies in the SrTiO$_3$ substrate of the Nd$_{0.8}$Sr$_{0.2}$NiO$_2$/SrTiO$_3$ heterostructure is negligible. Moreover, Ren et al. [104] recently reported the growth of superconducting Pr$_{0.8}$Sr$_{0.2}$NiO$_2$ thin films onto a non-SrTiO$_3$ substrate (LaAlO$_3$)$_{0.3}$(Sr$_2$AlTaO$_6$)$_{0.7}$ (LSAT), which further suggests that even if there is an inevitable little amount of oxygen vacancies in the SrTiO$_3$ substrates of superconducting Nd$_{0.8}$Sr$_{0.2}$NiO$_2$/SrTiO$_3$ heterostructures due to over reduction, they are not critical for the emergence of the superconductivity in nickelates.

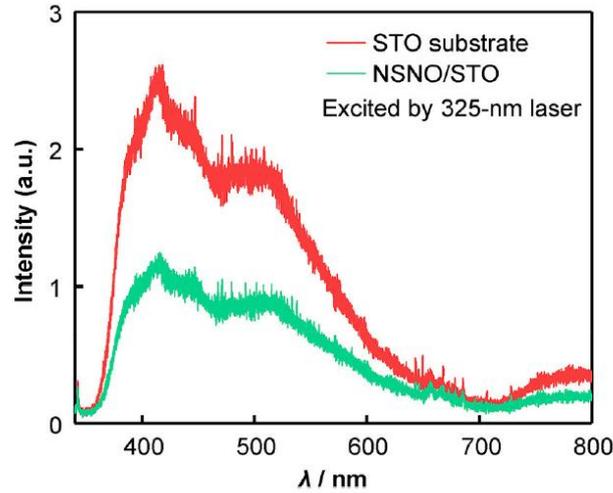

**FIGURE 10** Photoluminescence of a $Nd_{0.8}Sr_{0.2}NiO_2$/$SrTiO_3$ heterostructure and a bare $SrTiO_3$ substrate exited by a 325 nm laser. Reproduced with permission from Ref. [107]. Copyright 2021, Springer Nature.

## 2.5 A key puzzle

Unlike Cu- and Fe-based superconductors for which the superconductivity was firstly discovered in their bulk compounds, a key puzzle in this emerging field is that the superconductivity has been absent in bulk nickelates so far. This has largely limited the development of the emergent infinite-layer Ni-based superconductors as many characterization techniques are more accessible for bulk materials such as neutron diffraction, nuclear magnetic resonance, and angle-resolved photoemission spectroscopy that is capable of direct mapping of electronic band structures.

For example, in 2020, Li et al. [124] reported the absence of superconductivity in polycrystalline bulk $Nd_{0.8}Sr_{0.2}NiO_2$ compounds. Using a three-step method including chemical reduction, they successfully synthesized bulk $Nd_{0.8}Sr_{0.2}NiO_2$ with the same doping concentration as that of Ref. [26]. Fig. 11a illustrates the chemical reduction process that transforms $Nd_{1-x}Sr_xNiO_3$ to infinite-layer $Nd_{1-x}Sr_xNiO_2$. Powder X-ray diffraction shown in Fig. 11b confirms the formation of the phase.

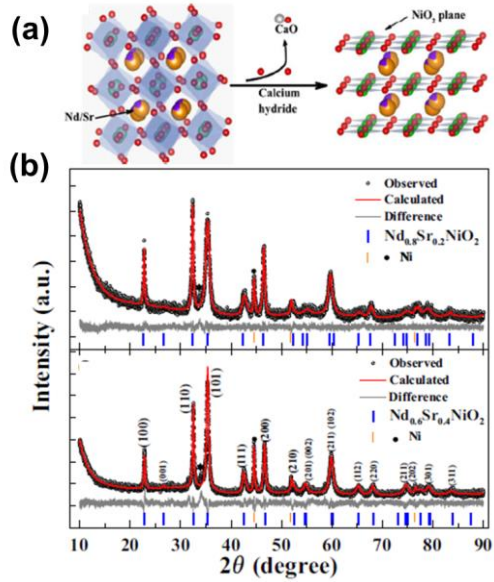

**FIGURE 11** (a) Schematic of chemical reduction from $Nd_{1-x}Sr_xNiO_3$ to infinite-layer $Nd_{1-x}Sr_xNiO_2$. (b) Powder X-ray diffraction pattern of $Nd_{1-x}Sr_xNiO_2$ ($x$ = 0.2 & 0.4). Reproduced with permission from Ref. [124]. Copyright 2020, Springer Nature.

Fig. 12a shows the resistivity as a function of temperature for the bulk $Nd_{1-x}Sr_xNiO_2$, which exhibits insulating behavior without any sign of superconductivity. Furthermore, high-pressure resistivity measurements were conducted to examine the possible superconductivity. Fig. 12b presents temperature-dependent resistivity at different pressures. One can see that even at a high pressure of 50.2 GPa, superconductivity still could not be found in bulk $Nd_{1-x}Sr_xNiO_2$. The authors suggested that interface effects uniquely existing in thin-film samples may be vital for the emergence of superconductivity in infinite-layer nickelates.

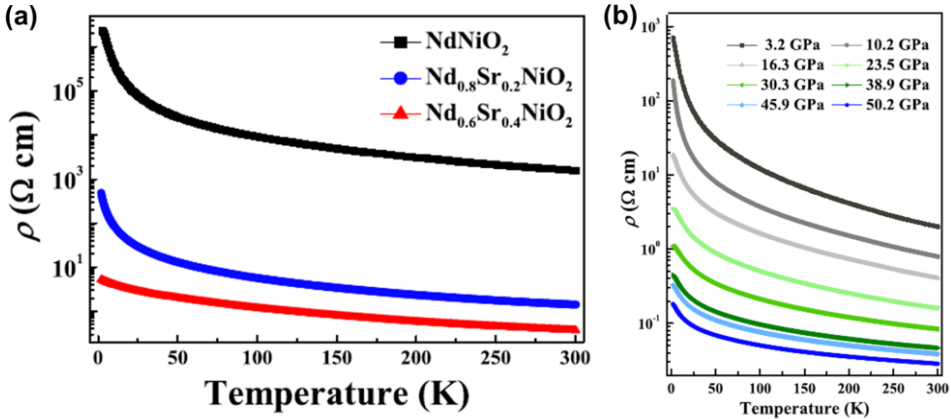

**FIGURE 12** (a) Temperature-dependent resistivity of polycrystalline bulk $NdNiO_2$ and $Nd_{1-x}Sr_xNiO_2$ ($x$ =0.2 & 0.4). (b) Temperature-dependent resistivity of the polycrystalline bulk $Nd_{1-x}Sr_xNiO_2$ at high pressures. Reproduced with permission from Ref. [124]. Copyright 2020, Springer Nature.

Soon later, Wang et al. [125] reported the similar absence of superconductivity in bulk polycrystalline $Nd_{1-x}Sr_xNiO_2$. They employed a sample fabrication technique combining the sol-gel combustion and high-pressure annealing to synthesize $Nd_{1-x}Sr_xNiO_3$ compounds, and then $Nd_{1-x}Sr_xNiO_2$ samples were converted from $Nd_{1-x}Sr_xNiO_3$ via chemical reduction reactions. The X-ray diffraction patterns of $Nd_{1-x}Sr_xNiO_3$ and $Nd_{1-x}Sr_xNiO_2$ shown in Fig. 13a fit well with the simulated results for the two phases. Electrical transport measurements display no hints of superconductivity in these samples (Fig. 13b).

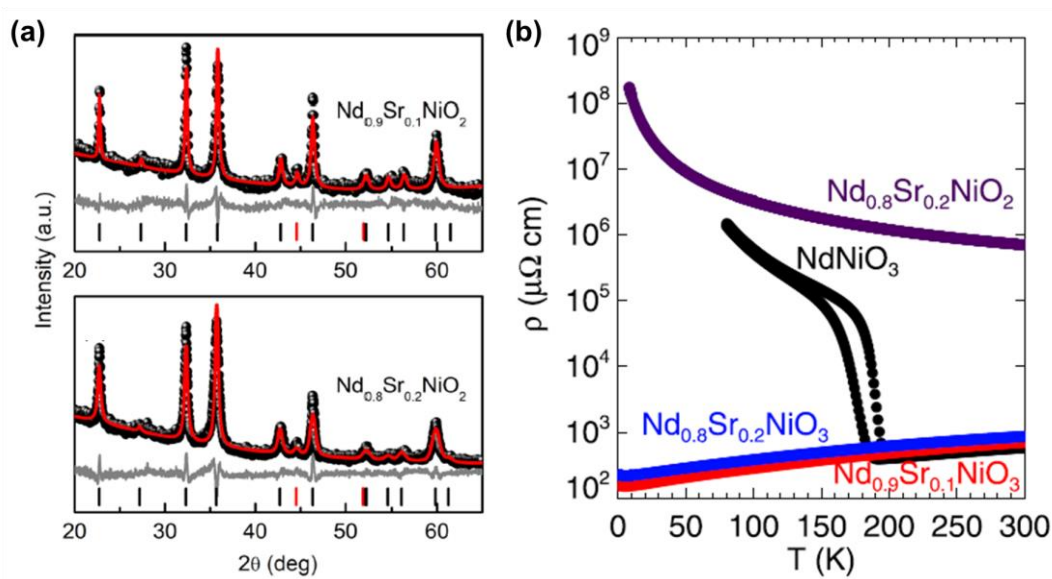

**FIGURE 13** (a) Powder X-ray diffraction patterns of bulk $Nd_{0.9}Sr_{0.1}NiO_2$ and $Nd_{0.8}Sr_{0.2}NiO_2$. (b) Temperature dependence of resistivity of bulk polycrystalline $NdNiO_2$, $Nd_{0.9}Sr_{0.1}NiO_2$ and $Nd_{0.8}Sr_{0.2}NiO_2$. Reproduced with permission from Ref. [125]. Copyright 2020. American Physical Society.

Up to now, more bulk infinite-layer nickelate compounds with other rare earth elements or other dopants have been successfully synthesized. For example, in 2020, He et al. [126] demonstrated the successful preparation of bulk polycrystalline infinite-layer $Sm_{0.8}Sr_{0.2}NiO_2$. In 2021, Puphal et al. [127] reported the synthesis of high-quality single-crystal infinite-layer $La_{1-x}Ca_xNiO_{2+\delta}$. Unfortunately, no superconductivity has been achieved yet in these compounds as well. These experimental results may suggest the vital role of interface effects in this type of thin-film heterostructures similar as that in other oxide [122,123] and two-dimensional-material-based heterostructures [128,129].

Meanwhile, theoretical researchers have definitely noticed this key puzzle and tried to seek possible explanations. Via density-functional theory, Si et al. [69] found out that the chemical reduction with CaH$_2$ may generate two different compounds $AB$O$_2$ and $AB$O$_2$H ($A$ is a rare-earth element, and $B$ is a transition-metal element). In particular, while $AB$O$_2$ is metallic, $AB$O$_2$H trends to be a Mott insulator. In an actual experiment, the mixture of these two phases could naturally emerge, which might prevent the macroscopic percolation of superconductivity. Moreover, it was proposed that the compressive strain in reduced infinite-layer thin films could avoid the hydrogen intercalation so that the metallic $AB$O$_2$ phase is energetically favored, which may thus assist in achieving superconductivity.

## 3. Electrical transport properties of nickelates

### 3.1 Critical current density

The critical current density ($J_c$) of a superconductor represents the maximal capability for it to carrying a current without resistance, which is a fundamental parameter for a new type of superconductors. Once the carrying current density exceeds the critical value, in the type-I superconductors, the resulting magnetic field on the surface could break up the strong correlation of the Cooper pairs. Under the condition of type-II superconductors, the magnetic flux taking the form of vortex in the mix state experiences a net force and moves when the current passes through. The vortex motion can produce an electric field parallel to the current and a resistance called flux-flow resistance appears. Hence, superconductors will lose the characteristic of superconductivity and exhibit electrical resistance which causes energy dissipation. In general, the critical current density is associated with the properties of samples themselves, such as grain boundary, porosity, impurity distribution, and purity of superconducting phase.

In the first experimental work of superconducting Nd$_{0.8}$Sr$_{0.2}$NiO$_2$ [30], thin films of ~10 nm in thickness were fabricated on the TiO$_2$-terminated SrTiO$_3$ (001) substrates. The authors reported that the critical current density of superconducting $Nd_{0.8}Sr_{0.2}NiO_2$ thin films at 2 K is ~170 kA·cm$^{-2}$ (Fig. 14a) via *I-V* measurements, which is one or two orders of magnitude smaller than that of some typical superconductors, e.g. $YBa_2Cu_3O_{7-x}$ (~8×10$^6$ A·cm$^{-2}$) [130], Nb (~10$^6$ A·cm$^{-2}$) [131], $MgB_2$ (~3.5×10$^7$ A·cm$^{-2}$) [132] and Fe-based superconductors ( ~2×10$^6$ A·cm$^{-2}$) [133]. Osada *et al.* [91] synthesized a 12-nm-thickness $Pr_{0.8}Sr_{0.2}NiO_2$ thin film which is a counterpart of $Nd_{0.8}Sr_{0.2}NO_2$. They observed the similar superconductivity below 12 K and the critical current density at 2 K in this work possesses the same magnitude as the value previously found for $Nd_{0.8}Sr_{0.2}NiO_2$ (Fig. 14b). Later, our group [107] successfully fabricated 14-nm-thick $Nd_{0.8}Sr_{0.2}NiO_2$ thin films on $SrTiO_3$ (001) substrates and the critical current density was determined to be only ~4 kA·cm$^{-2}$ at 1.8 K via resistance versus current measurements (Fig. 14c). The large difference between previous reports and our results may come from different measurement methods and thus different measure standards for the critical current density, for which the onset of a measurable finite non-zero resistance marks the critical current density in our measurements while the previous studies record the current densities leading to the largest voltage jumps in the *I-V* measurements as the critical values. Nevertheless, in either way, the critical current density of the emerging infinite-layer Ni-based superconductors is relatively low, which may be an intrinsic nature for nanoscale thin-film superconductors due to interface scatterings and could be enhanced in bulk superconductors if later the superconductivity can be realized.

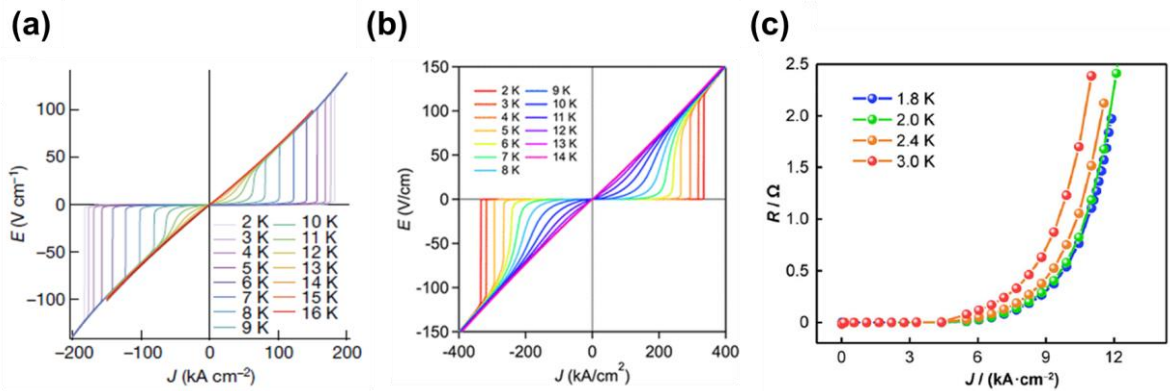

**FIGURE 14** (a) Electric field versus current density for a superconducting $Nd_{0.8}Sr_{0.2}NiO_2/SrTiO_3$ thin film from 2 to 16 K. Reproduced with permission from Ref. [30]. Copyright 2019, Springer Nature. (b) Electric field versus current density of a $Pr_{0.8}Sr_{0.2}NiO_2/SrTiO_3$ superconducting thin film for various temperatures. Reproduced with permission from Ref. [91]. Copyright 2020, American Chemical Society. (c) Current density-dependent resistance for a $Nd_{0.8}Sr_{0.2}NiO_2/SrTiO_3$ heterostructure at low temperatures. Reproduced with permission from Ref. [107]. Copyright 2021, Springer Nature.

**3.2 In-plane isotropic transport**

The lack of superconductivity in bulk nickelates implies that the novel phenomenon may be a surface or an interface effect via strain or interfacial electronic reconstruction. Since the interfaces between films and substrates could play a key role, the vicinal steps on the surfaces of $TiO_2$-terminated $SrTiO_3$ substrates due to miscut angles could influence the superconductivity. Accordingly, our group [107] has made linear four-probe electrical contacts with an Al wire bonder to conduct electrical measurements on a superconducting $Nd_{0.8}Sr_{0.2}NiO_2$ thin film along two orthorhombic in-plane edges (Fig. 15a). As shown in Fig. 15b, the normal-state resistivity exhibits anisotropic behavior for different measurement directions. The room-temperature resistivity for the direction perpendicular to the step of the $SrTiO_3$ substrate is ~7.7% larger than that for another in-plane orthorhombic direction, which indicates that the discrepancy of resistivity is associated with the electron scattering at the surface steps. Nevertheless, the superconducting transition temperature is almost the same for the two directions. One is ~9.5 K and the other is ~10 K for the onset transition temperature. It suggests that distinct resistivity of the normal state induced by the surface steps has only a negligible effect on the superconducting state of $Nd_{0.8}Sr_{0.2}NiO_2/SrTiO_3$ thin-film heterostructure. Thus, the in-plane isotropic superconductivity could be a common feature for the thin-film nickelate superconductivity.

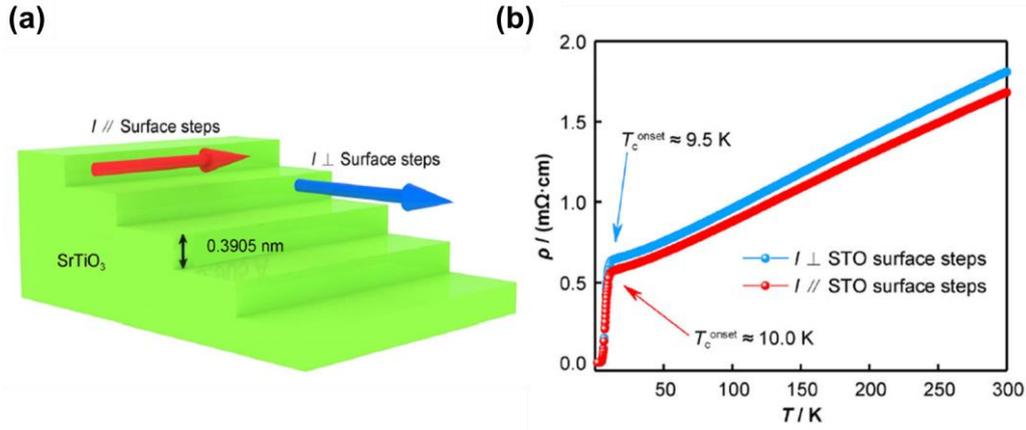

**FIGURE 15** (a) Schematic of the measurement geometry with in-plane orthometric directions. (b) Resistivity versus temperature for a $Nd_{0.8}Sr_{0.2}NiO_2$/$SrTiO_3$ heterostructure measured with the current applied along different in-plane directions. Reproduced with permission from Ref. [107]. Copyright 2021, Springer Nature.

### 3.3 Phase diagram

A phase diagram for a superconducting material system is fundamentally important as it can guide many relevant studies. Studies about doping-dependent superconducting phase diagrams are critical to understand the physics of nickelate superconductivity. It is commonly considered that cuprate superconductors are complicated material systems, characterized by a doping-dependent superconducting dome in a complex phase diagram, which seems to be true for nickelate superconductors as well. Recently, the experimental phase diagram of $Nd_{1-x}Sr_xNiO_2$ infinite-layer thin films was revealed by Li *et al.* from Stanford University [90]. As shown in Fig. 16a, a superconducting dome spanning $0.125 < x < 0.25$ is found, which is similar to cuprates. However, nickelates are weakly insulating in both underdoped and overdoped sides of the dome, which is different from cuprate superconductivity bounded by an insulator for underdoping and a metal for overdoping. Additionally, a small $T_c$ dip for $x = 0.2$ is observed. It is similar to the behavior of the dome in $La_{2-x}M_xCuO_4$ ($M$ = Sr or Ba) at $x = 1/8$, which is attributed to the presence of the stripe order. Shortly thereafter, Zeng et al. [102] from National University of Singapore also described the phase diagram of $Nd_{1-x}Sr_xNiO_2$ infinite-layer thin films (Fig. 16b). The doping-dependent superconducting dome extends between $x = 0.12$ and 0.235 with a small $T_c$ dip at $x = 0.18$, consistent with that in the report of Li et al. [90].

Furthermore, the later study from Stanford University [92] investigated the phase diagram of Pr$_{1-x}$Sr$_x$NiO$_2$ infinite-layer thin films (Fig. 16c). They found a superconducting dome of $0.12 < x < 0.28$ with a maximum $T_c$ of 14 K at $x = 0.18$, accompanied by weakly insulating behavior on both sides similar to Nd$_{1-x}$Sr$_x$NiO$_2$. However, in contrast with Nd$_{1-x}$Sr$_x$NiO$_2$, there is a broader superconducting doping range and no $T_c$ suppression is observed for superconducting Pr$_{1-x}$Sr$_x$NiO$_2$ thin films. These differences were ascribed to the strain field or chemical pressure induced by cation size disorder effects, which can affect the ground states of the system, and nickelates are more sensitive to the delicate balance between the local chemical environment and electronic correlations. Overall, the above works on superconducting nickelate phase diagrams have developed an in-depth understanding of nickelate superconductivity and provided the crucial experimental data for the synthesis of optimal nickelate superconductors.

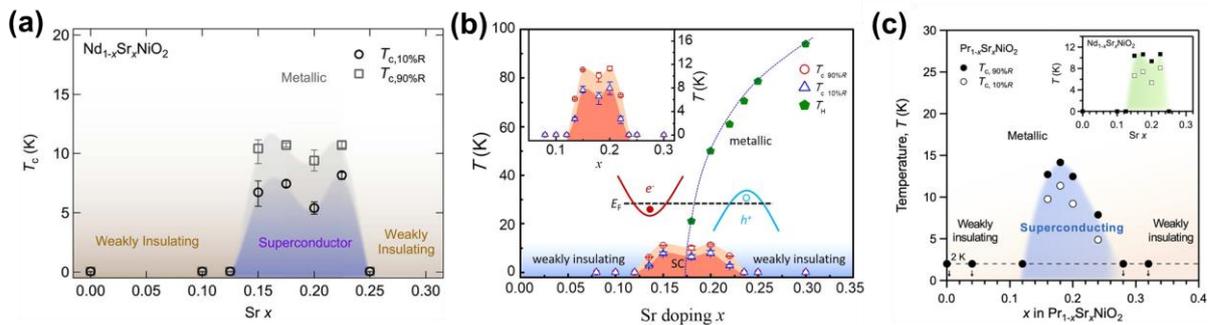

**FIGURE 16** (a) Phase diagram of Nd$_{1-x}$Sr$_x$NiO$_2$ infinite-layer thin films obtained by the Stanford University group. Reproduced with permission from Ref. [90]. Copyright 2020, American Physical Society. (b) Phase diagram of Nd$_{1-x}$Sr$_x$NiO$_2$ infinite-layer thin films obtained by the National University of Singapore group. Reproduced with permission from Ref. [102]. Copyright 2020, American Physical Society. (c) Phase diagram of Pr$_{1-x}$Sr$_x$NiO$_2$ infinite-layer thin films. Reproduced with permission from Ref. [92]. Copyright 2020, American Physical Society.

## 3.4 Upper critical field

The upper critical field ($\mu_0H_{c2}$) is another key parameter of type-II superconductors. Experimental measurements of the $\mu_0H_{c2}$ of the emerging infinite-layer nickelates Nd$_{1-x}$Sr$_x$NiO$_2$ are essential to investigate the fundamental pair breaking mechanisms and the possible anisotropy. As a layered superconductor, it is taken granted that (001)-oriented Nd$_{1-x}$Sr$_x$NiO$_2$ possesses anisotropic superconducting behavior such as different transition temperatures and $\mu_0H_{c2}$ for the in-plane and out-of-plane directions. Since the superconductivity is only realized in thin films, all the studies regarding the superconducting anisotropy focus on the measurements and fitting of the $\mu_0H_{c2}$. The researchers from Stanford University have conducted temperature-dependent resistivity measurements of an ~10-nm-thick $Nd_{0.775}Sr_{0.225}NiO_2$ film under various magnetic fields applied along the *c*-axis and *ab* plane and collected the temperature and field values at which the resistivity reaches 50% of the normal state [95]. Surprisingly, despite of the layered crystal structure, it was found that the $\mu_0H_{c2}$ of the $Nd_{0.775}Sr_{0.225}NiO_2$ film is isotropic and dominated by the spin response rather than orbital effects (Fig. 17a). Meanwhile, Xiang et al. from Hai-Hu Wen's group at Nanjing University studied the low-temperature magnetic transport properties of a 6-nm-thick $Nd_{0.8}Sr_{0.2}NiO_2$ film. They obtained $\mu_0H_{c2}$ from the temperature-dependent resistivity curves measured at different magnetic fields, using the criteria of 95% and 98% of the normal state resistivity [100]. However, different from the conclusion drawn by Stanford University, the group of Nanjing University declared a small anisotropic ratio ($\mu_0H_{c2}^{//ab}/\mu_0H_{c2}^{//c}$) of the measured $\mu_0H_{c2}$ located in the range of 1.2 to 3 (Fig. 17b). Besides, they deduced that the Pauli paramagnetic pair breaking mechanism plays a dominant role, which is consistent with the Stanford's conclusion.

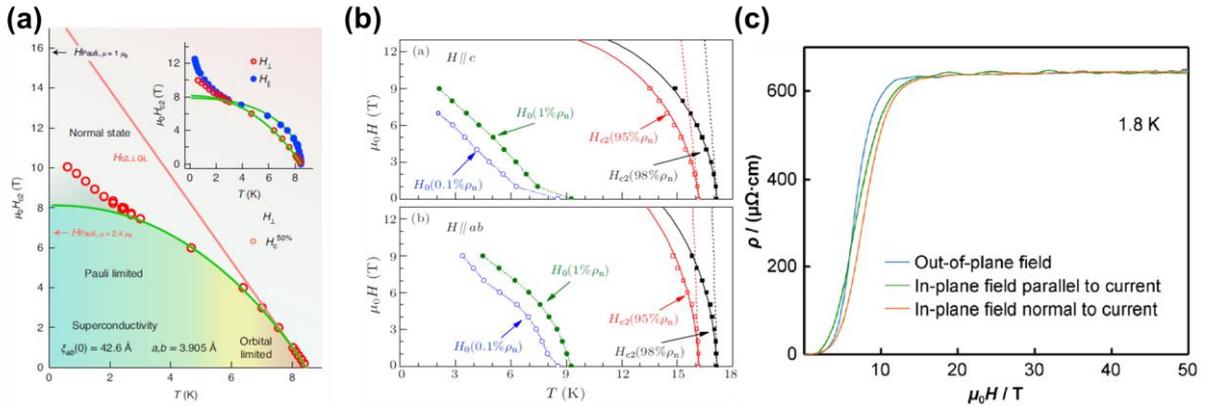

**FIGURE 17** (a) Temperature-dependent $\mu_0H_{c2}$ of an ~10-nm-thick $Nd_{0.775}Sr_{0.225}NiO_2$ film plotted by the group of Stanford University using the 50% criterion. Open and filled circles are collected under magnetic fields along the *c* axis and *ab* plane, respectively. Reproduced with permission from Ref. [95]. Copyright 2020, Springer Nature. (b) Temperature-dependent $\mu_0H_{c2}$ of a 6-nm-thick $Nd_{0.8}Sr_{0.2}NiO_2$ film sketched by the group of Nanjing University using the 95% and 98% criteria, respectively. Reproduced with permission from Ref. [100]. Copyright 2021, IOP Publishing. (c) Field-dependent resistivity of a 14-nm-thick $Nd_{0.8}Sr_{0.2}NiO_2$ film for three measurement geometries at 1.8 K. Reproduced with permission from Ref. [107]. Copyright 2021, Springer Nature.

To further clarify the anisotropy of the $\mu_0H_{c2}$, high-quality samples, lower temperature and higher magnetic fields are needed. Supported by the Wuhan National High Magnetic Field Center, our group performed systematic pulsed high-field magnetotransport measurements up to 50 T for a $Nd_{0.8}Sr_{0.2}NiO_2$ films with a thickness of 14 nm [107]. The results show that the $\mu_0H_{c2}$ at 1.8 K reaches ~11.5 T when the magnetic field is applied along the $c$ axis, which is generally consistent with the zero-temperature $\mu_0H_{c2}$ ~ 11.4 T obtained from static-field measurements. Nevertheless, the $\mu_0H_{c2}$ measured at a field along the $ab$ plane at 1.8 K is ~14.5 T, larger than that at a field along $c$ axis, which clearly indicates the anisotropy of the superconducting behavior (Fig. 17c). It is worth noting that when the magnetic field is applied along the $ab$ plane either parallelly or perpendicularly to the current, the field-dependent resistivity curves are slightly different, suggesting that the measurement geometry also has an impact on the conclusion.

Recently, Wang et al. from Jinguang Chen's group of Chinese Academy of Science reported the effect of pressure on the $\mu_0H_{c2}$ of the superconduct $Pr_{0.82}Sr_{0.18}NiO_2$ films [103]. They used the criteria of 50% of the normal state and Ginzburg-Landau formula to fit the zero-temperature $\mu_0H_{c2}$ and found that the zero-temperature $\mu_0H_{c2}$ exhibits a non-monotonic evolution with varied pressure ranging from 0 to 12.1 GPa. But when they plotted the normalized $\mu_0H_{c2}(T)/\mu_0H_{c2}(0)$ versus the normalized superconducting transition temperature $T_c^{mid}(\mu_0H)/T_c^{mid}(0)$, where $T_c^{mid}(\mu_0H)$ obtained from the temperature where the resistivity reaches 50% of the normal state at different magnetic field, they found that all the data at various pressures almost fall onto one single straight line, indicating that the main pair breaking mechanism related to $\mu_0H_{c2}$ barely changes under pressures.

Until now, superconducting $Nd_{1-x}Sr_xNiO_2$ samples are still scarce and the observation of superconductivity in bulk $Nd_{1-x}Sr_xNiO_2$ has not been achieved yet. Considering the possible large number of defects and mixed phases in thin films, it is reasonable for different groups to draw contradictory conclusions regarding the anisotropy of $\mu_0H_{c2}$ for these Ni-based superconducting thin films. The debate about whether the $\mu_0H_{c2}$ is anisotropic will probably continue for a while until high-quality superconducting bulk $Nd_{1-x}Sr_xNiO_2$ single crystals are available.

### 3.5 Multiband transport

Soon after the report of the first nickelate superconductor, a large number of fundamental works centered on electronic structures of $Nd_{1-x}Sr_xNiO_2$ were carried out. Experimentally, the groups from Stanford University and National University of Singapore systematically measured the temperature- and Sr-doping-dependent Hall coefficients ($R_H$) of the infinite-layer $Nd_{1-x}Sr_xNiO_2$ films [30,90,102]. All of them observed that the sign of $R_H$ changes from negative to positive as the temperature decreases and/or as the doping level increases, indicating that both electron and hole pockets are involved in the Fermi surfaces of the $Nd_{1-x}Sr_xNiO_2$ superconductors and the multiband-transport feature (Fig. 18a-b). Subsequently, it was found that $R_H$ of the newly discovered $Pr_{1-x}Sr_xNiO_2$ [92], $La_{1-x}Sr_xNiO_2$ [94] and $La_{1-x}Ca_xNiO_2$ [101] families also exhibit the sign change behavior similar to that of $Nd_{1-x}Sr_xNiO_2$, which further emphasizes the importance of electronic structure and electron filling level for the superconductivity.

Generally, the Hall resistance ($R_{xy}$) in a multiband regime varies nonlinearly versus the magnetic field. Especially in a two-band model, the density and mobility of electrons and holes can be deduced by fitting the nonlinear curves. However, in all the experiments mentioned above, $R_{xy}$ shows almost a perfectly linear feature under magnetic fields less than 9 T. Using a pulsed high-field up to 50 T, our group collected [107] the Hall voltage of the normal state of a $Nd_{0.8}Sr_{0.2}NiO_2$ film at different temperatures and observed the sign change of $R_H$ and nonlinearity of $R_{xy}$, providing direct transport evidence for the multiband electronic structure of Ni-based superconductors (Fig. 18c).

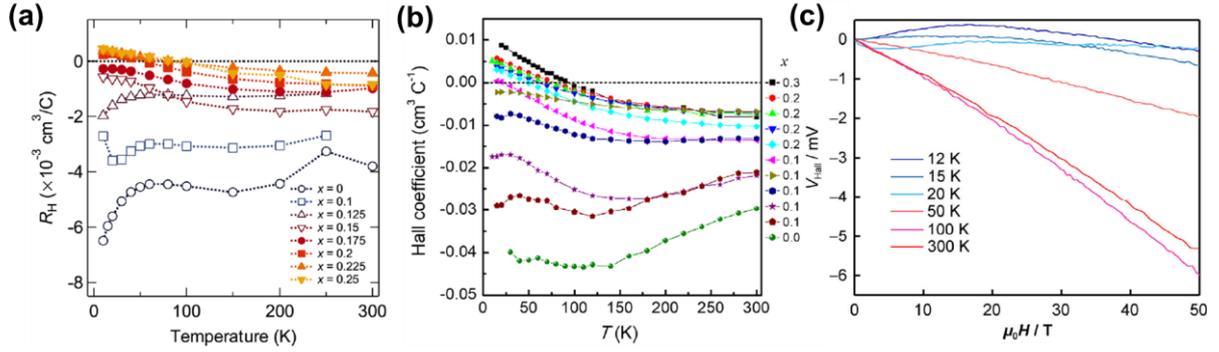

**FIGURE 18** (a) & (b) Temperature-dependent Hall coefficient $R_H$ for various Sr-doping levels obtained by the groups of Stanford University and National University of Singapore, respectively. Reproduced with permission from Ref. [90,102]. Copyright 2020, American Physical Society. (c) Hall voltage versus magnetic field up to 50 T at different temperatures. Reproduced with permission from Ref. [107]. Copyright 2021, Springer Nature.

In addition to the electrical transport evidence, based on an angstrom-size scanning transmission electron microscopy probe, Goodge et al. from Cornell University utilized locally resolved electron energy-loss spectroscopy to detect the electronic structures of parent perovskite NdNiO$_3$ and infinite-layer compounds, successfully establishing the Mott-Hubbard character of Nd$_{1-x}$Sr$_x$NiO$_2$ [97]. The data analysis of O-$K$, Ni-$L_{2,3}$ and Nd-$N_{2,3}$ edges evinces that the hole doping not only acts on O $2p$ but also on the Ni $3d$ and Nd $5d$ bands, which offers further evidence for the multiband electronic structure of the superconducting nickelates.

### 3.6 Are $f$ electrons essential for the superconductivity?

In Nd$_{1-x}$Sr$_x$NiO$_2$, the $4f$ orbits of the rare-earth element Nd hybridizes with Ni $3d$ orbits, and that naturally raises a question: whether the itinerant $f$ electrons are essential for the emergent superconductivity, reminiscent of the pivotal roles of $f$ electrons in heavy-fermion superconducting systems. Although relevant theoretical studies have revealed the important role of the hybridization between the $4f$ and $3d$ orbits of Nd$_{1-x}$Sr$_x$NiO$_2$ that can meanwhile regulates the magnetic state [44,52,87], the experimental verification of this key physical could be much more straightforward, i.e., whether one can realize superconductivity in an infinite-layer nickelate compound with less or eventually zero $f$ electrons, for example, replacing Nd with lower-$Z$ rare-earth elements or even La that does not possess an $f$ electron.

It turns out that while replacing Nd with Pr that possesses one less *f* electrons, infinite-layer Pr$_{1-x}$Sr$_x$NiO$_2$ thin films fabricated onto SrTiO$_3$ substrates by Harold Hwang's group (Fig. 19a) can indeed be superconducting [91,92]. Moreover, Xingjiang Zhou's group achieved [104] the superconductive Pr$_{1-x}$Sr$_x$NiO$_2$ films on both LSAT and SrTiO$_3$ substrates (Fig. 19b), which excludes the SrTiO$_3$-related substrate and interfacial effects.

Furthermore, replacing Nd with a zero-*f*-electron element such as La, both the Stanford University group and the National University of Singapore group have successfully obtained superconductivity (Fig. 19c-d) in either LaSrNiO$_2$ [94] or LaCaNiO$_2$ thin films [101]. Therefore, from what have been achieved experimentally, *f* electrons are not essential for superconductivity in the emergent infinite-layer nickelates despite of a number of previous theoretical studies stating the important roles of *f* electrons.

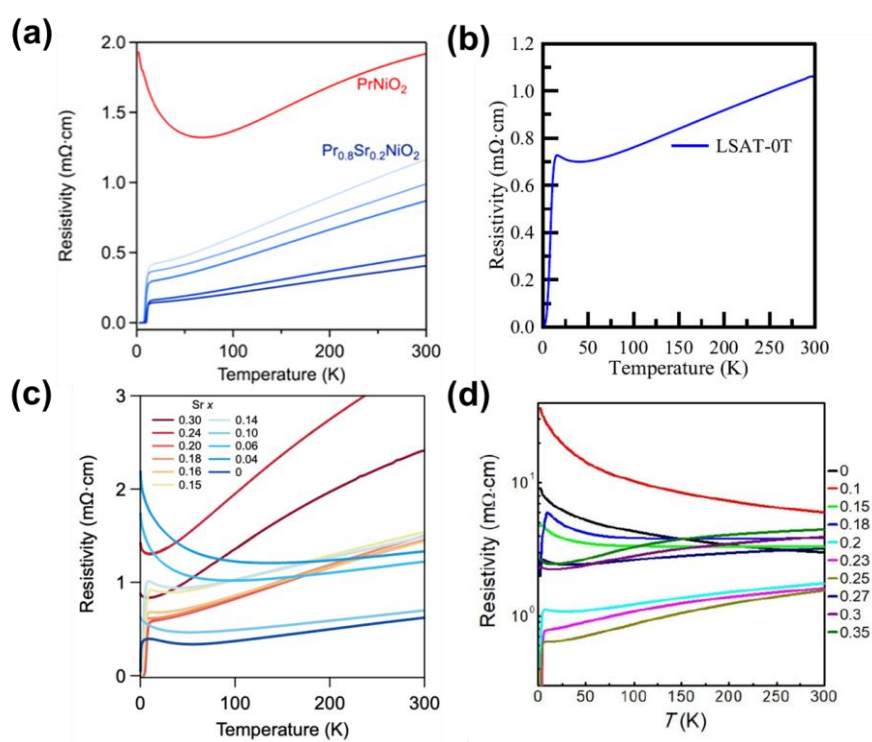

**FIGURE 19** (a) Temperature-dependent resistivity of Pr$_{0.8}$Sr$_{0.2}$NiO$_2$/SrTiO$_3$ films with different thickness ranging from 5.3 to 12 nm obtained by Harold Hwang's group. Reproduced with permission from Ref. [91]. Copyright 2020, American Chemical Society. (b) Temperature-dependent resistivity of a Pr$_{0.8}$Sr$_{0.2}$NiO$_2$/LSAT film obtained by Xingjiang Zhou's group. Reproduced with permission from Ref. [104]. (c) Temperature-dependent resistivity of La$_{1-x}$Sr$_x$NiO$_2$/SrTiO$_3$ thin films with various Sr doping obtained by Harold Hwang's group. Reproduced with permission from Ref. [94]. Copyright 2021, Wiley-VCH. (d) Temperature-dependent resistivity of La$_{1-x}$Ca$_x$NiO$_2$/SrTiO$_3$ films with different Ca doping obtained by Ariando's group. Reproduced with permission from Ref. [101].

## 3.7 Electric-field modulation of the superconductivity

Due to the strongly corelated nature of oxide materials, a slight change of the carrier density of oxide superconductors could intensively alter their transport behavior in both normal and superconducting states, which is rather intriguing and can serve as the central principle of superconducting field-effect transistors. Thus, it is of great interest to explore the electrostatic effect in this brand-new Ni-based superconducting system.

In April 2021, our group [107] studied the electrostatic carrier modulation of the superconducting transition of $Nd_{0.8}Sr_{0.2}NiO_2$/$SrTiO_3$ heterostructures. Via changing the polarity of the gating electric fields ($E_G$) perpendicularly applied onto the $SrTiO_3$ substrate, electrons/holes of ~$10^{13}$ cm$^{-2}$ under a small electric field of few kV/cm can be injected into the $Nd_{0.8}Sr_{0.2}NiO_2$ layer at low temperatures. As shown in Fig. 20, a positive $E_G$ of +3 kV/cm that brings electrons to the conducting $Nd_{0.8}Sr_{0.2}NiO_2$ channel slightly decreases $T_C$, while a negative $E_G$ of −4 kV cm$^{-1}$ that supplies holes subtly increases $T_c$. The electric-field modulation is rather similar to the electric-field tuning of another exemplary superconducting oxide, $YBa_2Cu_3O_7$ [134]. In addition, the normal state resistance is practically unchanged under $E_G$. Since the free-hole density of $Nd_{0.8}Sr_{0.2}NiO_2$ around its $T_c$ is considerably high (~$10^{16}$ cm$^{-2}$) compared to those of the injected carriers, it explains that the modulation effect is negligible for the normal state. In order to achieve effective carrier tuning in this system, more effective carrier injection approaches such as ferroelectric gating [135-137] and organic ionic gating [138,139] may be needed.

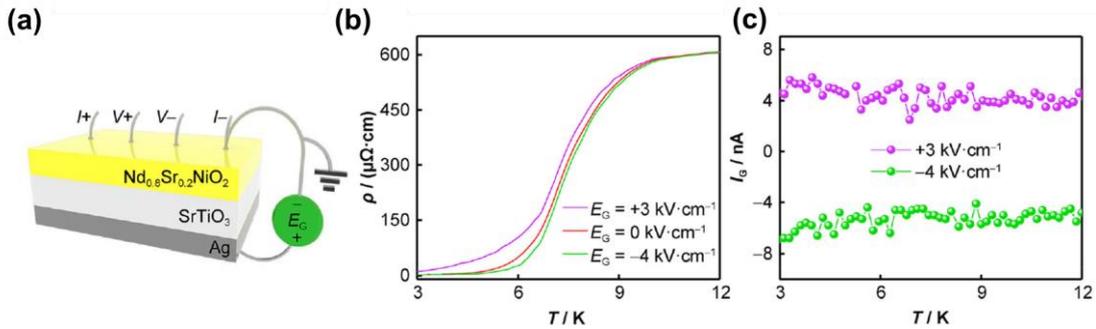

FIGURE 20 (a) Schematic of the measurement geometry of the electric-field modulation experiment. (b) Superconducting transitions for different gating electric fields ($E_G$). (c) Perpendicular gating current versus temperature under different $E_G$. Reproduced with permission from Ref. [107]. Copyright 2021, Springer Nature.

## 3.8 Pressure-modulated superconductivity with possible $T_c$ above 30 K

Superconductivity with higher transition temperatures has been always a holy grail for scientists to pursue. Static pressure serving as an independent non-thermal tuning parameter [140,141] has been frequently utilized as an effective handle to enhancing the $T_c$ of superconductivity. Recently, Wang *et al.* [104] fabricated superconducting infinite-layer $Pr_{0.82}Sr_{0.18}NiO_2$ thin films on $SrTiO_3$ substrates and an onset $T_c$ of ~18 K was achieved at ambient pressure. They further utilized a palm-type cubic anvil cell to exert hydrostatic pressure with glycerol as the liquid pressure transmitting medium. Although the normal-state and the residue resistivity increase abnormally with increasing pressure, the onset $T_c$ is raised to ~23.9 K at 4.6 GPa and is further enhanced to ~31 K at 12.1 GPa without leveling off (Fig. 21). The authors stated that the in-plane lattice constants of nickelate thin films are locked by the $SrTiO_3$ substrates whereas the *c* axis contracts under a positive pressure effect. It was indicated that the onset $T_c$ of the nickelate thin films reported from other groups till now seems to be inversely proportional to the constants of *c*-axis, which could be a key aspect for their enhanced $T_c$ under pressure. Besides, the increase of effective mass $m^*$ of charge carriers or the stronger electron correlations induced by the positive pressure could have important contributions to the enhancement of the $T_c$ as well.

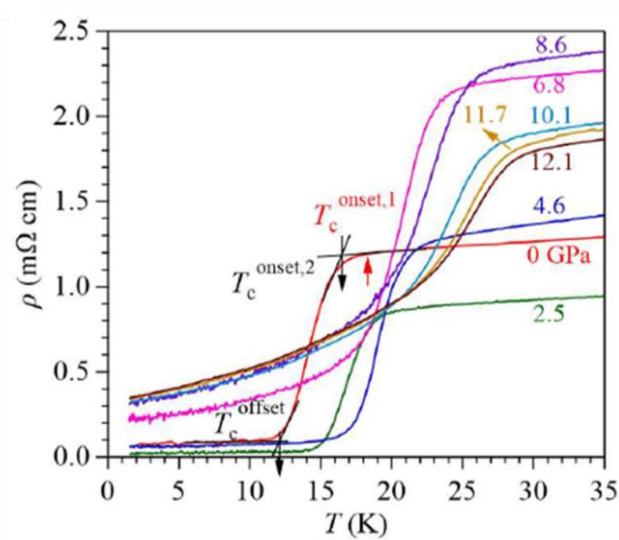

**FIGURE 21** Temperature-dependent resistivity for a $Pr_{0.82}Sr_{0.18}NiO_2$/$SrTiO_3$ thin film under various hydrostatic pressures ranging from 0 to 12.1 GPa. Reproduced with permission from Ref. [104]. Copyright 2021.

In order to realize the ultralow-energy-loss transportation, the $T_c$ of the superconductors is better to exceed 40 K that is generally called the McMillan limit [142]. One of the important challenges for the superconducting nickelate thin films is to enhance the superconducting transition temperature further. Higher pressure experiments beyond 12.1 GPa could be carried out on the nickelate thin films later and it is expected that the $T_c$ could be further raised. Moreover, the theoretical work for the evolution of electronic structures under pressure is necessary which could yield a better view on the electronic correlations in the superconducting nickelate thin films.

## 4. Magnetic properties

Due to the similarities in the crystal and electronic structures of layered nickelates with the well-known high-$T_c$ cuprates, a large number of studies have focused on the magnetic order of nickelates to explore new unconventional superconductors. As early as around 2000, magnetization measurements and neutron powder diffraction have been conducted in bulk $LaNiO_2$ [20] and $NdNiO_2$ [22]. However, no long-range antiferromagnetic order was found in these materials, possibly because the samples' low quality and the hindrance from ferromagnetic nickel impurities. In 2010, by combining neutron powder diffraction, nuclear magnetic resonance and density functional theory (DFT), Poltavets et al. [143] disclosed an antiferromagnetic transition at 105 K in a layered compound $La_4Ni_3O_8$ with a mixture of $Ni^{1+}$ and $Ni^{2+}$. To certain extent, these early explorations implied the complications of magnetic order in the layered nickelates.

Later in 2019, the breakthrough of superconducting $Nd_{0.8}Sr_{0.2}NiO_2$ thin films discovered by Li et al. stimulated substantial theoretical and experimental efforts to investigate the magnetic order of rare-earth nickelates and confirm that whether spin fluctuations could mediate superconductivity in $Nd_{1-x}Sr_xNiO_2$ superconductors similar to cuprates. Although Li et al.

claimed that magnetic order or spin fluctuations are absent (or considerably diminished) in the bulk compounds of nickelate superconductors based on the previous results of the magnetism of bulk LaNiO$_2$ and NdNiO$_2$ mentioned above, new DFT calculations and experimental measurements have found some trails of antiferromagnetism [38,47,53,56,61,62,64,84,86]. Theoretically, first-principles calculations for electronic and magnetic structures of undoped parent compounds $R$NiO$_2$ ($R$ = rare earth) have been launched. As a result, a phase transition with an estimated transition temperature of ~70-90 K from a paramagnet to a weak antiferromagnet [62] and a metallic antiferromagnetic ground state [53] for NdNiO$_2$ were proposed. On the premise that $R$-Ni coupling is suppressed, an antiferromagnetic insulator state has also been put forward [38]. For parent bulk compound NdNiO$_2$, besides the studies of antiferromagnetic order, calculations about antiferromagnetic exchange coupling $J$ have also been carried out by Katukuri et al. [56] and Wan et al. [86] and values of 77 meV and 82 meV were obtained, respectively. Moreover, Wan *et al.* emphasized that the Nd-5$d$ states and doping do not affect the exchange interactions $J$ significantly. On the other hand, some studies paid attention to other layered $d^9$ nickelates. Specifically, Nomura *et al.* found a significant $J$ as large as ~100 meV [64]; Fan et al. Theoretically investigated Ruddlesden–Popper-based nickelates Ln$_4$Ni$_3$O$_8$ (Ln = La, Nd) and found the existing antiferromagnetism in these materials [47]. These theoretical results intensively suggest that the antiferromagnetic order or spin fluctuations may be an intrinsic property and could be rather important for the superconductivity of Nd$_{1-x}$Sr$_x$NiO$_2$.

Experimentally, Raman scattering studies have provided direct evidence for the strong spin fluctuations in bulk NdNiO$_2$ and derived an antiferromagnetic exchange energy $J$ of ~25 meV [144]. Nuclear magnetic resonance experiments have been performed to confirm a considerable value of the antiferromagnetic exchange $J$ in bulk LaNiO$_2$ [145]. By resonant inelastic X-Ray scattering, Lu *et al.* observed a magnetic excitation with a bandwidth of about 200 meV and a substantial *J* of ~63.6 meV for superconducting nickelate thin films [93], while Lin et al. observed a *J* of 69(4) meV via resonant inelastic X-Ray scattering for the reduced trilayer nickelates $R_4Ni_3O_8$ (where $R$ = La, Pr) [146]. For trilayer bulk non-superconducting nickelates $Pr_4Ni_3O_8$/$La_4Ni_3O_8$, by powder neutron diffraction, magnetization measurements and muon spin rotation, complicated spin-glass behavior from 2 to 300 K and short-range correlated regions below ~70 K were suggested [147].

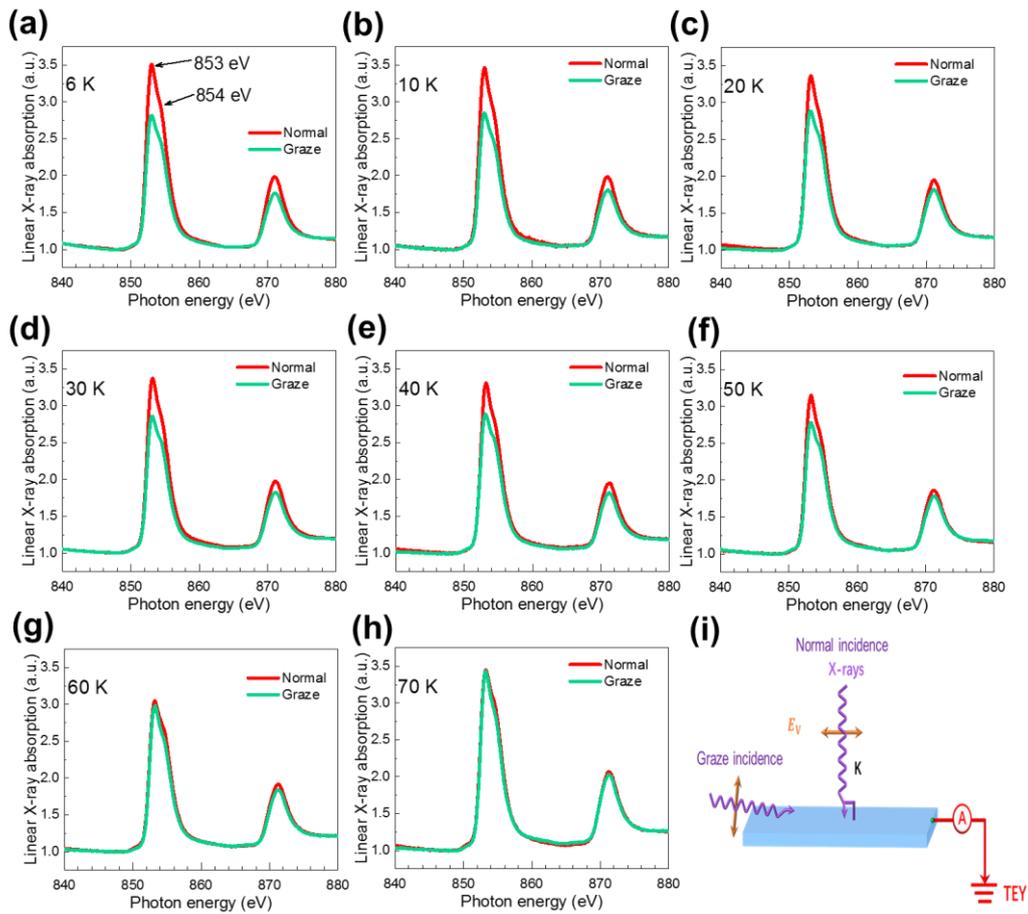

**FIGURE 22** (a)-(h) X-ray magnetic linear dichroism of a 20-nm-thick superconducting $Nd_{0.8}Sr_{0.2}NiO_2$/$SrTiO_3$ film. (i) Schematic of the measurement geometry. Reproduced with permission from Ref. [106]. Copyright 2021, Wiley-VCH.

More recently, the magnetic order of superconducting $Nd_{0.8}Sr_{0.2}NiO_2$ thin films have been systematically conducted by our group [106]. The X-ray magnetic linear dichroism data corroborate the long-range antiferromagnetic order of superconducting $Nd_{0.8}Sr_{0.2}NiO_2$ thin films (Fig. 22). Subsequently, by growing a 5-nm CoFe layer onto the superconducting thin films with thicknesses ranging from 8 to 40 nm, the exchange bias effect with different blocking temperatures was observed (Fig. 23a-b), which is in good agreement with the results of X-ray magnetic linear dichroism measurements. Based on our previous works about antiferromagnetic films [148-158], together with the X-ray magnetic circular dichroism spectroscopy measurements, the oxidation, impurity and other interface effects were excluded, hence confirming that the exchange bias originates from the interfacial magnetic coupling between the soft ferromagnetic CoFe layer and the antiferromagnetic order of the $Nd_{0.8}Sr_{0.2}NiO_2$ layer. It is worthy noticing that, for Sr-doped bulk non-superconducting $Nd_{0.85}Sr_{0.15}NiO_2$, a short-range antiferromagnetic order of magnetic moments below 40 K and antiferromagnetic spin fluctuations that continues to much higher temperatures (Fig. 23c) are recently evidenced by nuclear magnetic resonance measurements [159].

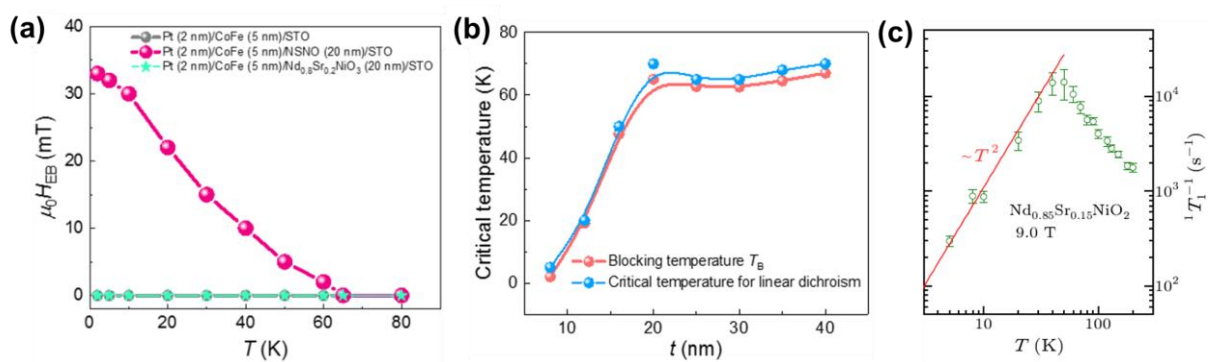

**FIGURE 23** (a) Temperature-dependent exchange bias $\mu_0 H_{EB}$ of a Pt (2 nm)/$Co_{90}Fe_{10}$ (5 nm)/$Nd_{0.8}Sr_{0.2}NiO_2$ (20 nm)/$SrTiO_3$ heterostructure and two reference samples Pt (2 nm)/$Co_{90}Fe_{10}$ (5 nm)/$SrTiO_3$ and Pt (2 nm)/$Co_{90}Fe_{10}$ (5 nm)/$Nd_{0.8}Sr_{0.2}NiO_3$ (20 nm)/$SrTiO_3$. (b) thickness-dependent critical temperatures for exchange bias and X-ray magnetic linear dichroism. Reproduced with permission from Ref. [106]. Copyright 2021, Wiley-VCH. (c) Temperature dependence of spin-lattice relaxation for bulk $Nd_{0.85}Sr_{0.15}NiO_2$ measured by nuclear magnetic resonance. The sharp peak represents an antiferromagnetic-type magnetic transition. Reproduced with permission from Ref. [159]. Copyright 2021, IOP Publishing.

Typically, the long-range antiferromagnetic order competes with superconductivity as the kinetic energy of free carriers could easily suppress the exchange interaction $J$. As a result, the superconductivity usually sets in when the antiferromagnetic order is killed by temperature, pressure, magnetic field, doping and other tuning parameters. However, for cuprates [160], Fe-based superconductors [161], heavy-fermion system [162], and quantum phase transition systems [163], antiferromagnetism could coexist with superconductivity under certain regions in their phase diagrams. Our experimental discovery on the coexistence of antiferromagnetism and superconductivity in $Nd_{0.8}Sr_{0.2}NiO_2$/$SrTiO_3$ thin films thus sheds light on the similarity between the emergent infinite-layer Ni-based superconductors and other well-known unconventional superconductors.

**5. Outlook**

In this final part, based on the abovementioned experimental aspects, we would like to then briefly and tentatively bring the most urgent experimental aspects out for this rapidly developing field:

(1) Firstly, how to possibly realize highly controllable and repeatable synthesis of the infinite-layer nickelate superconductor materials? The currently common chemical reduction approach via $CaH_2$ involves complex fabrication parameters and are difficult to be reproduce. Are there other possible means to stabilizing the 112-phase infinite-layer nickelates?

(2) Secondly, how to possibly understand the key puzzle of the absence of superconductivity in bulk infinite-layer nickelates? Interfacial effects induced by substrates in thin films such as strain, structural intermixing, electronic and orbital reconstructions? Tricks played by "invisible" H atoms that are difficult to be experimentally characterized? Nonuniformity in chemically reduced bulk samples with Ni aggregations? This is a fully open and the most urgent issue pending insightful experimental answer.

(3) Closely related to the second aspect, can the superconductivity be realized in bulk infinite-layer nickelate materials? Although it is challenging, once achieved, it would enable a significantly large range of in-depth physical characterizations of the emerging infinite-layer Ni-based superconductors and could lead to the blooming of relevant studies similar to Fe-based superconductors.

(4) Can the $T_c$ of the novel Ni-based superconductors be boosted, by chemical doping, pressure, electrostatic modulation, or even the isotope effect, to reach the McMillan limit 40 K or even higher? Higher $T_c$ for a new type of superconductors is an everlasting goal and sky shall be the limit in terms of what one can achieve for high-temperature superconductors.

(5) The magnetic phase diagram of the emergent infinite-layer superconducting nickelates is urgently needed to be constructed. With that, the correlation between magnetism (including spin fluctuations) and superconductivity could thus be clarified so that the most intriguing pairing mechanism for the exotic superconductivity may be revealed.

In conclusion, as an important component for materials science, the emergent infinite-layer Ni-based superconductivity has been well in its infant period and there is still a rather long way to go. More and more intensive efforts, especially at this stage experimental efforts, are needed to push this field forward steadily and rapidly.

**Acknowledgements**
X.Z., P.Q., and Z.F. contributed equally to this work. Z.L. acknowledges financial support from the National Natural Science Foundation of China (Nos. 52121001, 51822101, 51861135104 & 51771009).

**Data availability**
The data that support plots within this paper are available from the corresponding author upon reasonable request.